\documentclass[aps,prl,floatfix,twocolumn,reprint,amsmath,amssymb,superscriptaddress,UTF8]{revtex4-2}
\usepackage[english]{babel}

\usepackage{amsfonts}
\usepackage{mathrsfs}
\usepackage{amsmath}
\usepackage{color}
\usepackage{natbib}
\usepackage{graphicx}
\usepackage{bm}
\usepackage{amssymb}
\usepackage{xspace}
\usepackage{epstopdf}
\usepackage{dcolumn}
\usepackage{longtable}
\usepackage{array}
\usepackage{makecell}
\usepackage{tabulary}
\usepackage{multirow}
\usepackage{mhchem}
\usepackage{wrapfig}
\usepackage{pifont}  

\newcolumntype{C}[1]{>{\centering\arraybackslash}p{#1}}
\newcolumntype{L}[1]{>{\flushleft\arraybackslash}p{#1}}

\usepackage{multirow}
\usepackage{epsfig}
\usepackage[normalem]{ulem}
\usepackage{amsmath}
\usepackage{braket}
\usepackage{bbold}
\usepackage{natbib}

\usepackage[colorlinks=true, letterpaper=true, pdfstartview=FitV, linkcolor=blue, citecolor=blue, urlcolor=blue]{hyperref}

\makeatletter

\newcommand{\Rmnum}[1]{\expandafter\@slowromancap\romannumeral #1@}
\makeatother

\begin{document}

\title{Electric Hidden Unconventional Collinear Magnetism in 2D \(\mathcal{PT}\)-Symmetric Antiferromagnets}

\author{Mu Tian}
\thanks{These authors contributed equally to this work.}
\affiliation{Key Lab of advanced optoelectronic quantum architecture and measurement (MOE), Beijing Key Lab of Nanophotonics $\&$ Ultrafine Optoelectronic Systems, and School of Physics, Beijing Institute of Technology, Beijing 100081, China}

\author{Fanxian Pei}
\thanks{These authors contributed equally to this work.}
\affiliation{Key Lab of advanced optoelectronic quantum architecture and measurement (MOE), Beijing Key Lab of Nanophotonics $\&$ Ultrafine Optoelectronic Systems, and School of Physics, Beijing Institute of Technology, Beijing 100081, China}

\author{Zhi-Ming Yu}
\affiliation{Key Lab of advanced optoelectronic quantum architecture and measurement (MOE), Beijing Key Lab of Nanophotonics $\&$ Ultrafine Optoelectronic Systems, and School of Physics, Beijing Institute of Technology, Beijing 100081, China}
\affiliation{International Center for Quantum Materials, Beijing Institute of Technology, Zhuhai, 519000, China}

\author{Yugui Yao}
\affiliation{Key Lab of advanced optoelectronic quantum architecture and measurement (MOE), Beijing Key Lab of Nanophotonics $\&$ Ultrafine Optoelectronic Systems, and School of Physics, Beijing Institute of Technology, Beijing 100081, China}
\affiliation{International Center for Quantum Materials, Beijing Institute of Technology, Zhuhai, 519000, China}

\author{Run-Wu Zhang}
\email{zhangrunwu@bit.edu.cn}
\affiliation{Key Lab of advanced optoelectronic quantum architecture and measurement (MOE), Beijing Key Lab of Nanophotonics $\&$ Ultrafine Optoelectronic Systems, and School of Physics, Beijing Institute of Technology, Beijing 100081, China}
\affiliation{International Center for Quantum Materials, Beijing Institute of Technology, Zhuhai, 519000, China}

\date{\today}

\begin{abstract}
Electric hidden unconventional collinear magnetism, in which a perpendicular electric field (\(\mathcal{E}_z\)) lifts the \(\mathcal{PT}\)-enforced degeneracy and selectively releases the hidden crystallographic constraints in conventional 2D antiferromagnets to drive emergent magnetic phases, has remained largely unexplored. Here we establish a complete symmetry classification of all 88 collinear spin layer groups (cSLGs) compatible with conventional \(\mathcal{PT}\)-symmetric antiferromagnets, and show that the \(\mathcal{E}_z\) drives the system into three distinct classes of unconventional collinear magnetism: type-IV magnetism (Class-I, 23 cSLGs), altermagnetism (Class-II, 34 cSLGs), and fully compensated ferrimagnetism (Class-III, 31 cSLGs). We illustrate these categories using the most common and feasibly exfoliable 2D candidate materials. These include Sr$_2$CaCr$_2$P$_2$O$_9$ and Sr$_2$CaCr$_2$As$_2$O$_9$ (Class-I), BaMnGe and MnPS$_3$ (Class-II), and Sr$_3$Fe$_2$O$_5$Cl$_2$ and bilayer MnBi$_2$Te$_4$ (Class-III). We further demonstrate that the \(\mathcal{E}_z\) generates distinctly different spin-splitting fingerprints across the three classes, yielding distinct spin-dependent responses: the electric Hall effect in Class-I, spin-polarized transport in Class-II, and the electric reversible magneto-optical Kerr effect in Class-III. Our work not only closes a complete internal cycle from conventional to unconventional  collinear magnetism in 2D antiferromagnets, but also establishes a viable all-electrical compensated magnetism platform and control scheme for future spintronic applications.
\end{abstract}
\maketitle

\textit{\textcolor{blue}{Introduction.}}---The intimate connection between magnetic order and crystalline symmetry is a central theme in condensed matter physics for understanding and designing novel magnetic states~\cite{alexanderSymmetryConsiderationsDetermination1962,dimmockUseSymmetryDetermination1963,donnaySymmetryMagneticStructures1958,tokuraMagneticTopologicalInsulators2019,deng2020quantum,elcoroMagneticTopologicalQuantum2021,bernevigProgressProspectsMagnetic2022,zhangMagneticTopologicalMaterials2023,wuNodalSurfaceSemimetals2018a,PhysRevLett.128.066602,tangDiracFermionsAntiferromagnetic2016,kurebayashiMagnetismSymmetrySpin2022,liuSymmetryClassificationMagnetic2026}. The long-established paradigm holds that a system’s magnetic symmetry classification and electronic spin structure are constrained by the symmetry operations connecting its spin sublattices in the nonrelativistic collinear limit, an approach that has successfully categorized current crystal-symmetry-compensated magnets~\cite{bradley1972,yuanGiantMomentumdependentSpin2020b,yuanPredictionLowZCollinear2021,chenEnumerationRepresentationTheory2024f,jiangEnumerationSpinSpaceGroups2024c,xiaoSpinSpaceGroups2024b,liuSymmetryClassificationMagnetic2026}. Yet, what if this seemingly complete classification is merely part of a larger hidden landscape? What if symmetries beyond the primitive cell, concealed beneath the protection of these dominant operations, hold the key to emergent phases previously invisible to the conventional framework?

Among various magnetic systems~\cite{deng2020quantum,maMultifunctionalAntiferromagneticMaterials2021c,zhouManipulationAltermagneticOrder2025,sciadv.adn0479,yuQuantizedAnomalousHall2010a,chenIntrinsicMagneticTopological2019,zhanFloquetEngineeringNonequilibrium2023,zhanPerspectiveFloquetEngineering2024,smejkalConventionalFerromagnetismAntiferromagnetism2022f,smejkalEmergingResearchLandscape2022b,duanAntiferroelectricAltermagnetsAntiferroelectricity2025a,zhuAltermagneticProximityEffect2026,shaoNeelSpinCurrents2023a,zhangXtypeStackingCrosschain2025,chengSpinPumpingSpinTransfer2014a,danielsSpintransferTorqueInduced2015,jungwirthAntiferromagneticSpintronics2016b,Šmejkal2018,tangDiracFermionsAntiferromagnetic2016,watanabeMagneticParityViolation2024,zhangParityTimeSymmetryMagnetic2024,fengDiscoveryWeylNodal2019a,zhangNodalLineSpinGapless2020b,baiAnomalousHallEffect2025f,tianSymmetryClassificationAltermagnetism2026,mazinEditorialAltermagnetismNew2022a,DEGROOT199145,degrootNewClassMaterials1983,huHalfMetallicAntiferromagnetProspective2012,katsnelsonHalfmetallicFerromagnetsBand2008a,rimmlerNoncollinearAntiferromagneticSpintronics2025,vanleukenHalfMetallicAntiferromagnets1995a,luoSpinGroupSymmetry2026,hellenes2023p,brekke2024minimal,PhysRevLettweng,liStackingdependentElectronicTopological2025}, conventional \(\mathcal{PT}\)-symmetric antiferromagnets precisely embody this tension~\cite{deng2020quantum,chengSpinPumpingSpinTransfer2014a,danielsSpintransferTorqueInduced2015,jungwirthAntiferromagneticSpintronics2016b,Šmejkal2018,tangDiracFermionsAntiferromagnetic2016,watanabeMagneticParityViolation2024,zhangParityTimeSymmetryMagnetic2024}. These materials possess the combined symmetry of spatial inversion (\(\mathcal{P}\)) and time reversal (\(\mathcal{T}\)) in relativistic case, which protects their degenerate band structures and ensures zero net magnetization~\cite{tangDiracFermionsAntiferromagnetic2016,watanabeMagneticParityViolation2024,zhangParityTimeSymmetryMagnetic2024,Šmejkal2018}. Yet, it is this very protection that also masks the influence of other crystallographic operations on the electronic structure, leading to a common perception that conventional \(\mathcal{PT}\)-symmetric antiferromagnets form a single rigid class with very limited tunability~\cite{jungwirthAntiferromagneticSpintronics2016b,chengSpinPumpingSpinTransfer2014a,danielsSpintransferTorqueInduced2015}. In three dimensions, \(\mathcal{PT}\)-symmetric antiferromagnets have been extensively studied, but the effect of an out-of-plane electric field (\(\mathcal{E}_z\)) on breaking \(\mathcal{P}\) is often limited by sample thickness and electrostatic screening. In two dimensions, however, the situation changes dramatically and field control is emerging to serve as an efficient method for manipulating spin transport~\cite{matsukuraControlMagnetismElectric2015,kajiwaraTransmissionElectricalSignals2010,shenMagnetoelectricTorquePolar2026,xieManipulationTopologyElectric2025,yuElectricallySwitchableNonrelativistic2025a}. The reduced dimensionality not only enriches the possible magnetic ground states but also allows \(\mathcal{E}_z\) to efficiently penetrate the material and selectively break a subset of crystallographic operations while preserving others, thereby offering an unprecedented route to unveil hidden magnetic phases that are strictly forbidden in the pristine \(\mathcal{PT}\)-symmetric state~\cite{sun2025multifield,du2020berry,z69z-175r,qayyum2024pt,guo2024electric,wang2025electric}. 
Despite its great potential, a complete symmetry-based framework that systematically classifies how \(\mathcal{E}_z\) transforms conventional \(\mathcal{PT}\)-symmetric antiferromagnets  into unconventional collinear magnetic phases is still lacking. Previous classifications either rely on spin space groups and are limited to three dimensions, or do not consider the selective symmetry breaking by the electric field, leaving the \(\mathcal{E}_z\)-driven magnetic phase transitions and their full classification in two-dimensional (2D) \(\mathcal{PT}\)-symmetric antiferromagnets largely unexplored. 

\begin{figure*}[ht]
	\begin{center}  
		\includegraphics[width=1\textwidth]{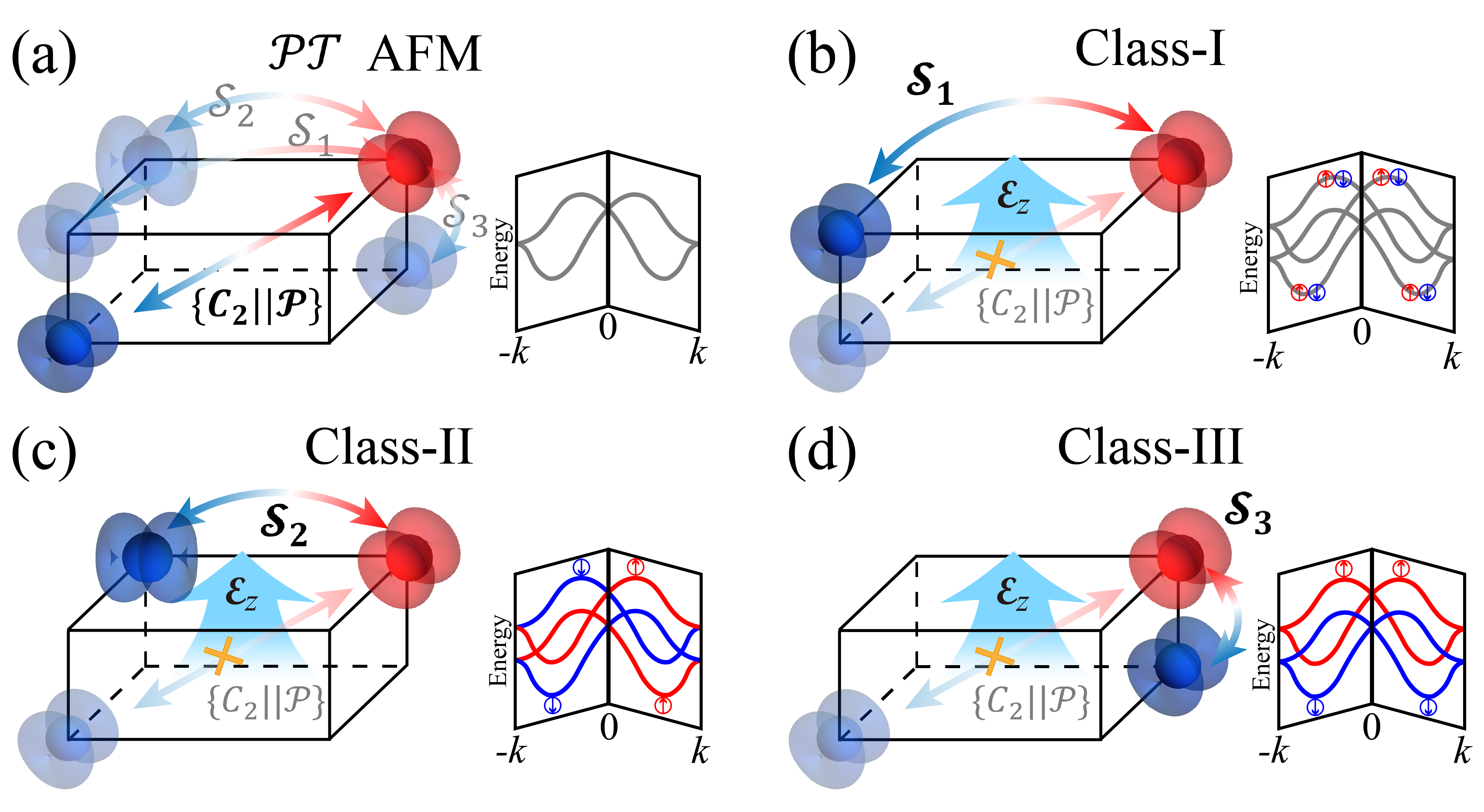}
		\caption{Schematic illustration of how a vertical electric field \(\mathcal{E}_z\) drives a conventional 2D \(\mathcal{PT}\)-symmetric antiferromagnet (a) into three distinct classes of unconventional collinear magnets: (b) type-IV magnet (Class-I), (c) altermagnet (Class-II), and (d) fully compensated ferrimagnet (Class-III). For each class, we show the schematic spin configurations of the two opposite sublattices and the symmetry operations \(\{C_2||\mathcal{P}\}\), \(\mathcal{S}_1\), \(\mathcal{S}_2\), and \(\mathcal{S}_3\) that connect them. The yellow cross marks the breaking of the \(\{C_2||\mathcal{P}\}\) symmetry under \(\mathcal{E}_z\). We also show the corresponding spin-resolved band structures for the three classes in the nonrelativistic limit. The gray bands represent the spin-degenerate states without \(\mathcal{E}_z\), while the red and blue bands display the opposite spin-splitting patterns induced by \(\mathcal{E}_z\).}\label{fig:1}  
	\end{center}
\end{figure*}

To fill this gap, we establish a complete symmetry classification of conventional 2D \(\mathcal{PT}\)-symmetric antiferromagnets within the collinear spin layer groups (cSLGs) framework~\cite{zhangSpinLayerGroups2026}, and show that an \(\mathcal{E}_z\) can drive them into three distinct classes of hidden unconventional collinear magnetism. According to the symmetry operations that survive the \(\mathcal{E}_z\)-induced selective symmetry breaking, the 88 cSLGs describing 2D \(\mathcal{PT}\)-symmetric antiferromagnetism fall into three categories: Class-I (electric type-IV magnetism, 23 cSLGs), Class-II (electric altermagnetism, 34 cSLGs), and Class-III (electric fully compensated ferrimagnetism, 31 cSLGs). To illustrate these categories, we select the most common or feasibly exfoliated 2D candidates, including Sr$_2$CaCr$_2$P$_2$O$_9$ and Sr$_2$CaCr$_2$As$_2$O$_9$ (Class-I), BaMnGe and MnPS$_3$ (Class-II), and Sr$_3$Fe$_2$O$_5$Cl$_2$ and bilayer MnBi$_2$Te$_4$ (Class-III). We demonstrate that \(\mathcal{E}_z\) efficiently lifts Kramers degeneracy and generates qualitatively different spin patterns, which manifest as distinct spin-polarized responses: Class-I systems can exhibit the electric Hall effect, Class-II systems can exhibit spin-dependent transport, and Class-III systems can exhibit the electric reversible magneto-optical Kerr effect (MOKE). These responses depend on the specific symmetry and electronic structure and are not universal to all systems within each class. Our work establishes a unified framework bridging conventional antiferromagnets and emerging magnetic phases, opening a viable route for low-power multifunctional spintronics based on 2D magnets.

\textit{\textcolor{blue}{Theoretical analysis.}}---The cSLGs theory provides the natural language for classifying 2D collinear magnets~\cite{zhangSpinLayerGroups2026}. Compared with spin space groups~\cite{chenEnumerationRepresentationTheory2024f,jiangEnumerationSpinSpaceGroups2024c,xiaoSpinSpaceGroups2024b,chenUnconventionalMagnonsCollinear2025b}, cSLGs more accurately capture the non-relativistic symmetries of 2D systems. Within this framework, conventional 2D $\mathcal{PT}$-symmetric antiferromagnetism is defined as collinear magnetism carrying $\{C_2||\mathcal{P}|\tau\}$ symmetry, which, in the relativistic limit, reduces to the combined $\mathcal{PT}$ (representing the combined operation of spatial inversion $\mathcal{P}$ and time reversal $\mathcal{T}$) symmetry and enforces Kramers spin degeneracy. Our previous cSLG classification of 2D magnets has already identified type-IV magnetism as a phase absent in three dimensions~\cite{baiAnomalousHallEffect2025f,tianSymmetryClassificationAltermagnetism2026}. Here, we generalize this approach to all 88 cSLGs compatible with conventional $\mathcal{PT}$-symmetric antiferromagnetism, which can be expressed as
$
	\mathcal{L}_\mathcal{P} = \{E||L\} + \{C_2||\mathcal{P }L\},
	\label{eq:Lp}
$
where $L$ is an index-two normal subgroup of $\mathcal{L}_\mathcal{P}$ consisting of spin-invariant operations.

Previous studies have frequently grouped conventional $\mathcal{PT}$-symmetric antiferromagnetism as a single class, as the $\{C_2||\mathcal{P}\}$-enforced spin degeneracy hides the influence of other symmetries. In 2D systems, however, an $\mathcal{E}_z$ breaks $\mathcal{P}$. This selective symmetry breaking divides all symmetry operations into two sets:
$
	\mathcal{O}_I = \{E,\; C_{nz},\; M_{\parallel}\},
$
which remain intact under $\mathcal{E}_z$, and
$
	\mathcal{O}_{II} = \{\mathcal{P},\; S_{nz},\; M_z,\; C_{2\parallel}\},
$
which are broken~\cite{cuiElectricHallEffect2025}. Here, $E$ represents identity element, $C_{nz}$ is an \textit{n}-fold in-plane rotation, and $M_{\parallel}$ denotes any mirror perpendicular to the plane; $S_{nz} = C_{nz}M_z$ denotes an improper rotation. Consequently, the symmetry operations in $\mathcal{P} L$ (after field application) that survive the field dictate the band structure. This yields three distinct scenarios (Fig.~\ref{fig:1}): (i) if $\exists \mathcal{S}_1 \in\mathcal{P} L,\mathcal{S}_1 = C_{2z}$, the spin degeneracy cannot be lifted by $\mathcal{E}_z$. In this case, the $\mathcal{E}_z$-applied system can then be regarded as an electric type-IV magnetism (Class-I); (ii) if $\exists\,\mathcal{S}_2\in
\mathcal{P}L,\mathcal{S}_2\in\mathcal{O}_I\setminus\{C_{2z}\}$, then after $\mathcal{E}_z$ is applied the system still possesses the $\bigl\{C_2|| \mathcal{S}_2\bigr\}$ symmetry, which can be regarded as electric altermagnetism (Class-II); (iii) if $\forall\mathcal{S}_3 \in \mathcal{P} L,\mathcal{S}_3\in \mathcal{O}_{II}$, then every operation that connects opposite spin sublattices will be destroyed by $\mathcal{E}_z$. The system is regarded as electric ferrimagnetism (Class-III). If the system remains semiconducting under \(\mathcal{E}_z\), it becomes an electrically induced fully compensated ferrimagnet~\cite{liuTwoDimensionalFullyCompensated2025a}.

Applying these criteria to all 88 cSLGs compatible with 2D conventional $\mathcal{PT}$-symmetric antiferromagnetism yields a complete classification: 23 belong to Class-I, 34 to Class-II, and 31 to Class-III (see Supplemental Material (SM)~\cite{SM}). Two representative examples illustrate the physical consequences. Bilayer MnBi$_2$Te$_4$, as a widely studied intrinsic magnetic topological insulator, is described by cSLG L.72.69.1.1, which falls into Class-III; an $\mathcal{E}_z$ thus induces ferromagnetic spin splitting in its originally degenerate bands. Monolayer MnPS$_3$, a layered antiferromagnetic semiconductor, belongs to cSLG L.71.67.1.1 (Class-II), and hence an applied $\mathcal{E}_z$ produces altermagnetic spin splitting. These examples demonstrate that the classification is not merely formal, and each class directly maps onto experimentally accessible electric-field responses in known 2D materials.

The above classification is not limited to specific compounds but applies universally to all 88 cSLGs describing 2D conventional 
$\mathcal{PT}$-symmetric antiferromagnetism. To explicitly demonstrate the distinct electronic responses of each class, we perform first-principles calculations on three representative materials, one for each class: monolayer Sr\(_2\)CaCr\(_2\)P\(_2\)O\(_{9}\) (Class-I), BaMnGe (Class-II) and Sr\(_3\)Fe\(_2\)O\(_5\)Cl\(_2\) (Class-III). Our classification thus unveils a purely electrical paradigm for 2D $\mathcal{PT}$-symmetric antiferromagnets: a simple static means $\mathcal{E}_z$ enables selective and reversible switching between three distinct unconventional magnetic phases, paving the way for multifunctional all-electrical spintronics.

\begin{figure}[t!]
	\centering 
	\includegraphics[width=0.5\textwidth]{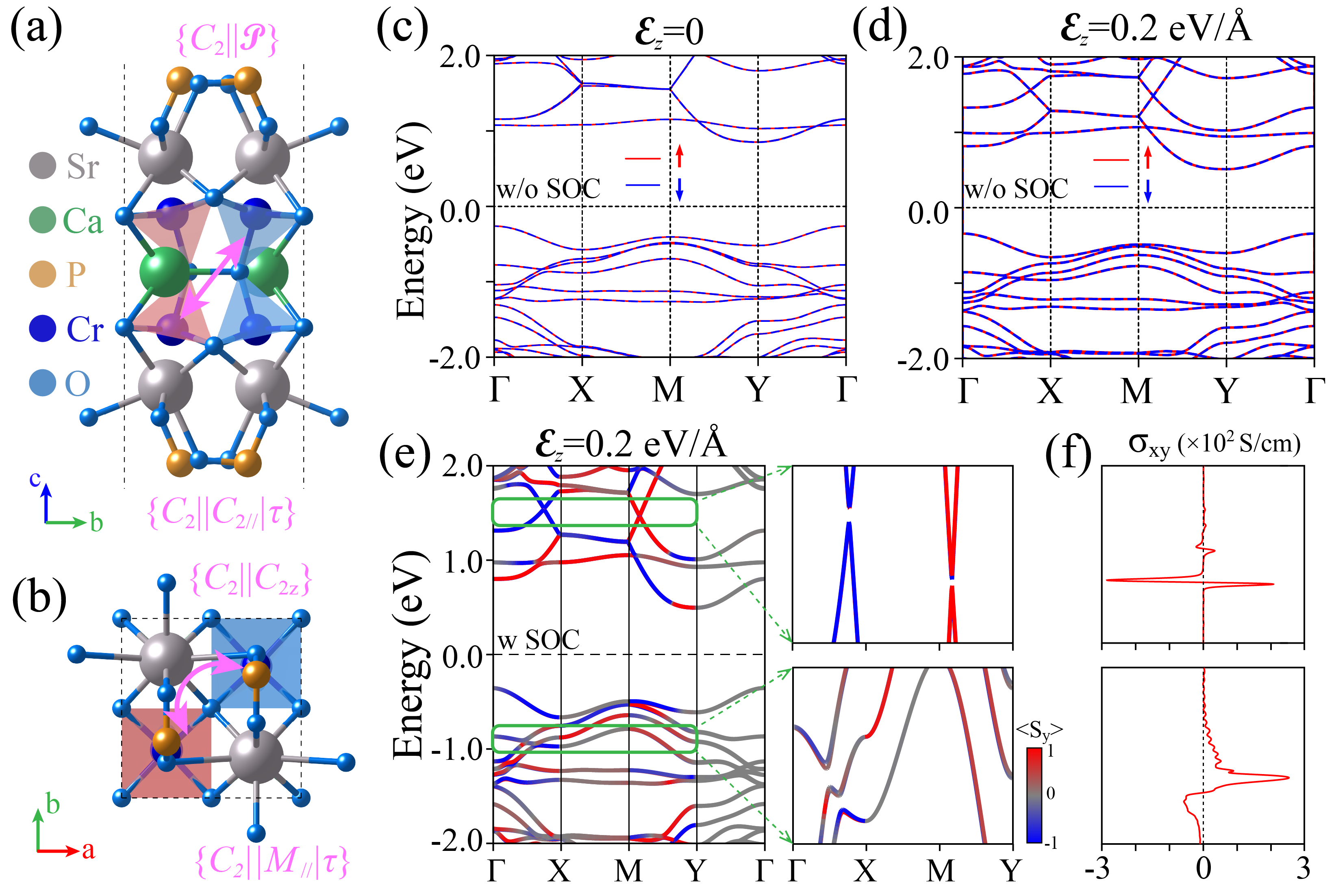}
	\caption{Crystal structure of monolayer \(\mathrm{Sr_2CaCr_2P_2O_9}\): (a) side and (b) top views. Spin-polarized band structures in the nonrelativistic limit (w/o SOC): (c) without and (d) with a vertical electric field \(\mathcal{E}_z\). The red solid and blue dashed lines represent spin-up and spin-down channels, respectively. In the presence of \(\mathcal{E}_z\), (e) \({S}_y\)-resolved relativistic (w SOC) band structure and (f) anomalous Hall conductivity \(\sigma_{xy}\), with the N\'eel vector oriented along the \([010]\) direction.} \label{fig2}  
\end{figure}

\textit{\textcolor{blue}{Class-I: electric type-IV magnetism.}}--Monolayer Sr\(_2\)CaCr\(_2\)X\(_2\)O\(_{9}\) (X=P, As) family serves as a potential representative, belonging to the electric type-IV cSLG L.41.31.1.1. Taking Sr\(_2\)CaCr\(_2\)P\(_2\)O\(_{9}\) as an example, its magnetic atoms occupy the 4\textit{k} Wyckoff position, and the two spin sublattices are related by $\{C_{2}||\mathcal{ P}\}$, $\{C_{2}|| C_{2z}\}$, 
$\{C_{2}|| C_{2\parallel}|\tau\}$, and 
$\{C_{2}|| M_{\parallel}|\tau\}$ [Figs.~\ref{fig2}(a) and \ref{fig2}(b)]. 
In the absence of $\mathcal{E}_z$, the spin-up and spin-down 
bands remain degenerate along the $\Gamma\text{-}X\text{-}M\text{-}Y\text{-}\Gamma$ path [Fig.~\ref{fig2}(c)]. This degeneracy follows from the symmetry constraint \(E(s, \mathbf{k}) = E(s, -\mathbf{k})\) enforced by \(\{ \bar{C}_2 || T \}\) for all collinear magnets~\cite{smejkalConventionalFerromagnetismAntiferromagnetism2022f,litvinSpinGroups1974,litvinSpinPointGroups1977a}; together with the symmetry relations above, it yields:
$$
\begin{aligned}
	\left\{\bar{C}_2 \| \mathcal{T}\right\}\left\{C_2 \| \mathcal{P}\right\} E(s, \mathbf{k}) & =E(-s, \mathbf{k}) , \\
	\left\{\bar{C}_2 \| \mathcal{T}\right\}\left\{C_2 \| C_{2z}\right\} E(s, \mathbf{k}) & =E(-s, \mathbf{k}) .
\end{aligned}
$$
Upon applying $\mathcal{E}_z$, the band dispersions 
are significantly modified, while the spin-up and spin-down bands remain 
degenerate because $\{C_{2}|| \mathcal{P}\}$ is broken whereas $\{C_{2}|| C_{2z}\}$ is preserved [Fig.~\ref{fig2}(d)]. In the nonrelativistic limit, this $\mathcal{E}_z$-unbreakable spin degeneracy identifies the system as Class-I.

When relativistic effects are taken into account, type-IV magnets often give rise to fascinating physical properties~\cite{baiAnomalousHallEffect2025f,tianSymmetryClassificationAltermagnetism2026}. In particular, after considering spin-orbit coupling (SOC), many physical phenomena may emerge in Class-I. For example, in monolayer Sr\(_2\)CaCr\(_2\)P\(_2\)O\(_{9}\) with the magnetic moment oriented along the [010] direction, the recently defined electric-field-induced Hall effect~\cite{cuiElectricHallEffect2025} can also appear. Moreover, the Hall conductivity exhibits a sign reversal when the electric field is reversed. Without $\mathcal{E}_z$, monolayer Sr\(_2\)CaCr\(_2\)P\(_2\)O\(_{9}\) will maintain Kramers degeneracy due to its \(\mathcal{PT}\) symmetry, and the Hall conductivity \(\sigma_{xy}\) is also constrained to be 0 (see details in SM\cite{SM}). When \(\mathcal{E}_z=0.2\,\mathrm{eV}/\text{\AA}\) is further applied, the band exhibits a subtle splitting [see Fig.~\ref{fig2}(e) and the enlarged view in the green box]. Consequently, the sum of the Berry curvatures contributed by the band-inversion points in the conduction band and the valence band is no longer zero, thereby giving rise to a significant Hall conductivity [Fig.~\ref{fig2}(f)]. When the electric-field direction is reversed (\(\mathcal{E}_z = \text{-}0.2\,\mathrm{eV}/\text{\AA}\)), the spin component \(S_y\) near the gaps also changes sign and the corresponding Hall conductivity reverses sign accordingly (see details in SM\cite{SM}). From a symmetry perspective, a Class-I under an applied \(\mathcal{E}_z\) essentially belongs to the recently discovered type-IV magnetism~\cite{baiAnomalousHallEffect2025f,tianSymmetryClassificationAltermagnetism2026}. Therefore, it may host many of the physical properties recently uncovered in this class of magnetism and holds great promise for applications.

	\begin{figure}[t!]
	\centering 
	\includegraphics[width=0.5\textwidth]{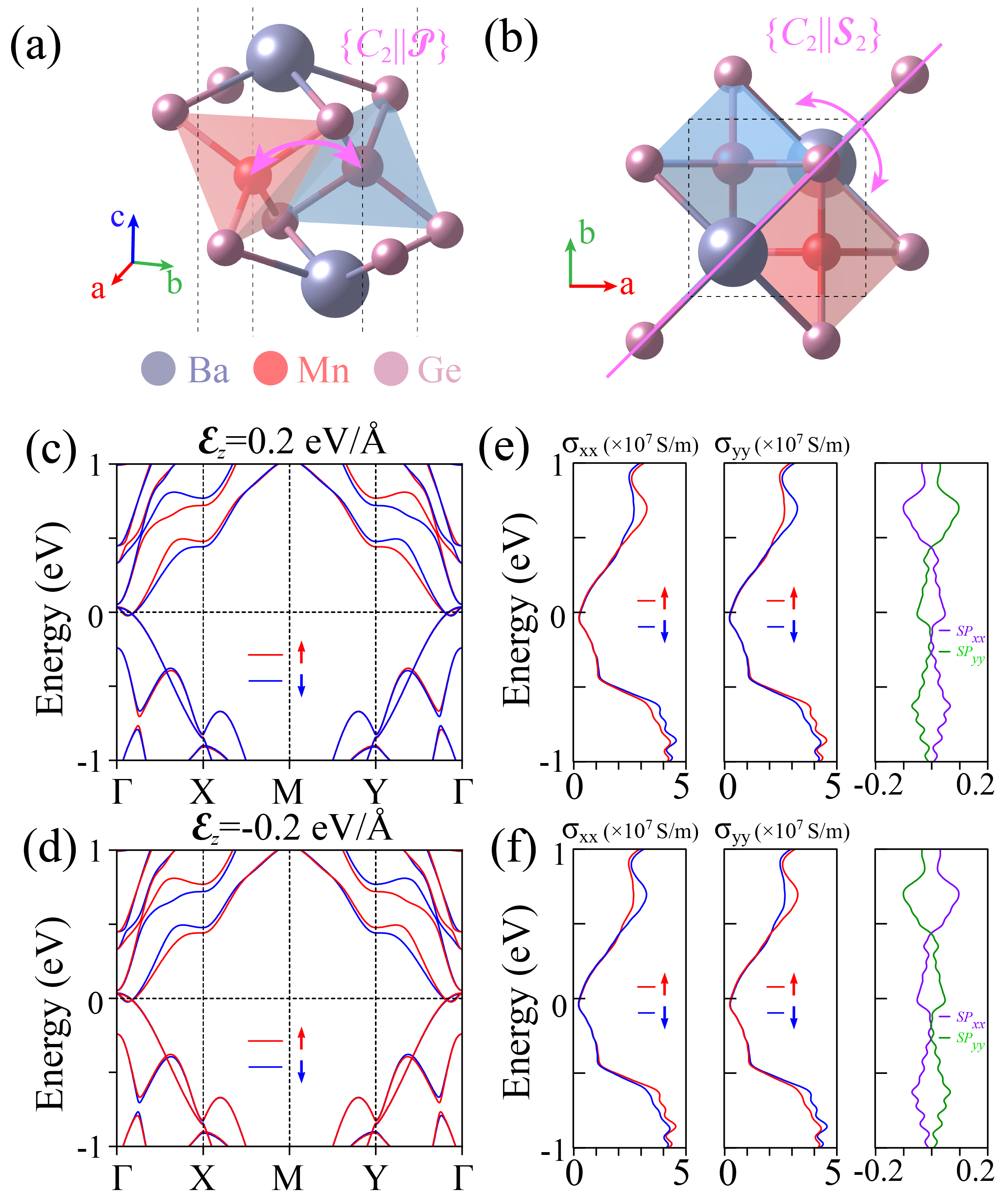}
	\caption{(a) Side and (b) top views of monolayer \(\mathrm{BaMnGe}\). Spin-resolved band structures obtained in the nonrelativistic limit (w/o SOC): with  a vertical electric field (c) $\mathcal{E}_{z}=0.2\,\mathrm{eV}/\text{\AA}$ and (d) $\mathcal{E}_{z}=\text{-}0.2\,\mathrm{eV}/\text{\AA}$. (e), (f) Corresponding conductivities \((\sigma_{xx}, \sigma_{yy})\) and corresponding spin polarizations \(({SP}_{xx}, {SP}_{yy})\) as functions of energy for (c) and (d).
	} \label{fig:3}  
\end{figure}

\textit{\textcolor{blue}{Class-II: electric altermagnetism.}}--Monolayer BaMnGe belongs to cSLG L.64.59.1.1, where the two spin sublattices are connected by the symmetry operations \(\{C_{2}|| \mathcal{P}\}\) and \(\{C_{2}|| \mathcal{S}_{2}\}\) [Figs.~\ref{fig:3}(a) and \ref{fig:3}(b)]. Here, \(\mathcal{S}_2=\{M_{\parallel}\tau,C_{4z}\tau,M_z\tau,C_{2\parallel}\tau\}\), with $\tau$ denoting a distinct sliding operation. In the absence of \(\mathcal{E}_z\), \(\{C_{2}|| \mathcal{P}\}\) protects the degeneracy between the spin channels (see details in SM\cite{SM}).

Upon applying an \(\mathcal{E}_z\), the symmetries \(\{C_{2}|| \mathcal{P}\}\), \(\{C_2||C_{2\parallel}\}\), and \(\{C_2||M_{z}\}\) are broken, whereas \(\{C_{2}|| M_{\parallel}\}\) and \(\{C_{2}|| C_{4z}\}\) are retained. The remaining symmetries impose directional constraints on the splitting, leading to an alternating, momentum-dependent spin splitting throughout the Brillouin zone [Figs.~\ref{fig:3}(c) and \ref{fig:3}(d)]. Such a symmetry-driven evolution is characteristic of the electric altermagnetic response discussed within the cSLG framework. 

Without \(\mathcal{E}_z\), the spin-up and spin-down contributions to both \(\sigma_{xx}\) and \(\sigma_{yy}\) are essentially identical, consistent with the spin-degenerate band structure  (see details in SM\cite{SM}). We define the spin polarization of conductivities as:
\begin{equation}
	SP_{\hat{n}\hat{n}}
	=
	\frac{
		\sigma^{\uparrow}_{\hat{n}\hat{n}}
		-
		\sigma^{\downarrow}_{\hat{n}\hat{n}}
	}{
		\sigma^{\uparrow}_{\hat{n}\hat{n}}
		+
		\sigma^{\downarrow}_{\hat{n}\hat{n}}
	}.
\end{equation} Accordingly, the corresponding spin polarizations \(SP_{xx}\) and \(SP_{yy}\) remain zero.  
A strikingly different behavior emerges under \(\mathcal{E}_z\) [Figs.~\ref{fig:3}(e) and \ref{fig:3}(f)]: the spin-up and spin-down longitudinal conductivities become clearly separated in both the \(x\) and \(y\) directions, accompanied by finite \(SP_{xx}\) and \(SP_{yy}\). This demonstrates that the electric field converts the initially spin-degenerate charge transport into spin-selective transport. Importantly, \(SP_{xx}\) and \(SP_{yy}\) exhibit distinct energy dependences and opposite tendencies, indicating that the spin polarization is strongly direction-dependent. Therefore, the system supports electrically tunable spin-polarized currents along orthogonal directions.

	\begin{figure}[t!]
	\centering 
	\includegraphics[width=0.5\textwidth]{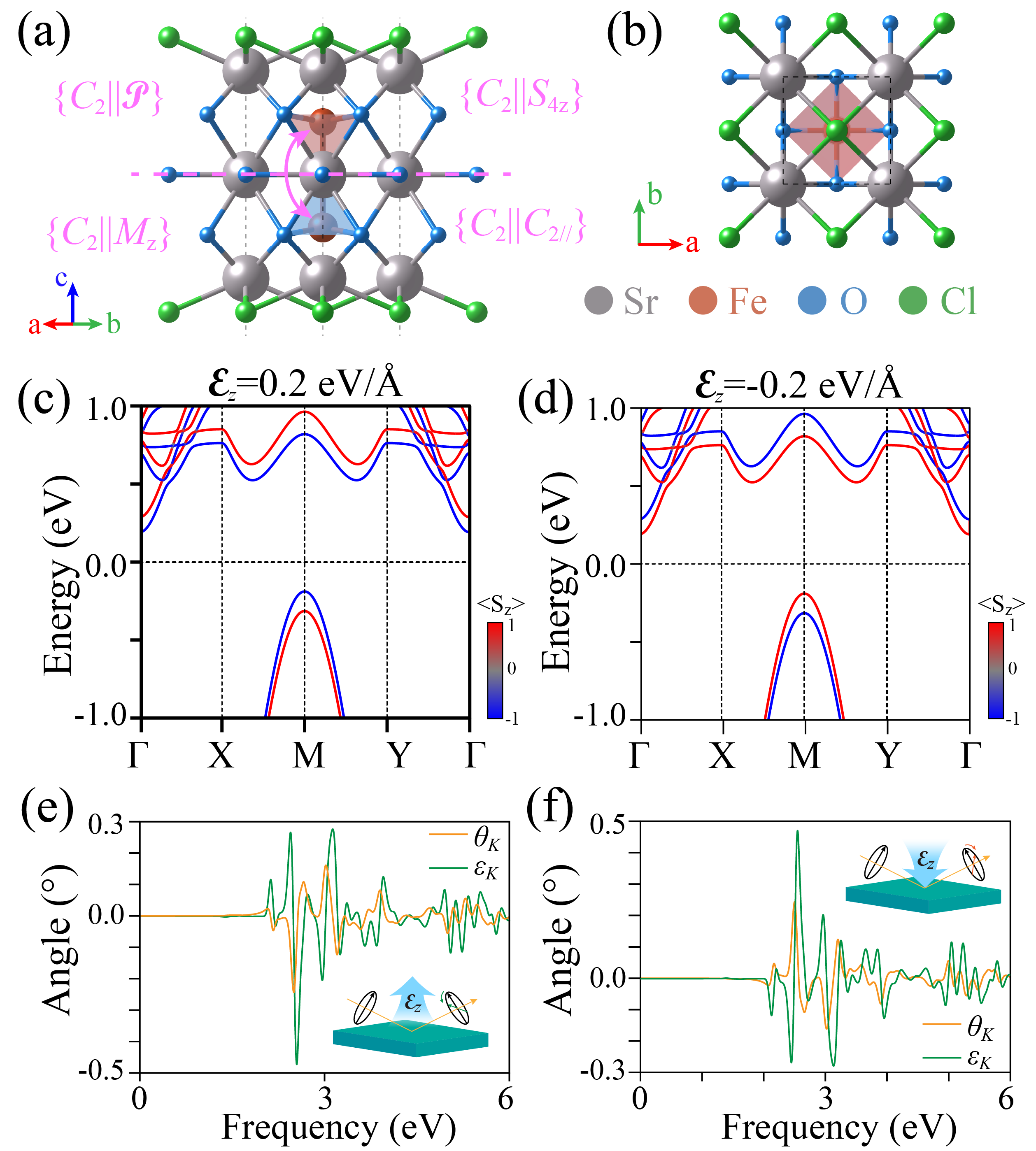}
	\caption{Crystal structure of monolayer Sr\(_3\)Fe\(_2\)O\(_5\)Cl\(_2\) in (a) side and (b) top views. (c, d) Spin-polarized band structures in relativistic case with the N\'eel vector oriented along the \([001]\) direction. Spin-resolved band structures with (c) positive and (d) negative \(\mathcal{E}_z\). (e), (f) Corresponding  Kerr rotation angle (\(\theta_{K}\)) and Kerr ellipticity (\(\varepsilon_{K}\)) as functions of frequency in (c) and (d).} \label{fig:4}  
\end{figure}
\textit{\textcolor{blue}{Class-III: electric fully compensated ferrimagnetism.}}--In the monolayer Sr\(_3\)Fe\(_2\)O\(_5\)Cl\(_2\) shown in Figs.~\ref{fig:4}(a) and ~\ref{fig:4}(b), the two spin sublattices are correlated by the symmetry operations \(\{C_{2}|| \mathcal{S}_3\}\), where \(\mathcal{S}_3=\{\mathcal{P},S_{4z},M_z,C_{2\parallel}\}\). In the absence of \(\mathcal{E}_z\), the symmetry \(\{C_2||\mathcal{P}\}\) protects the degeneracy between the two opposite-spin channels (see details in SM\cite{SM}).
Upon applying \(\mathcal{E}_z\), all \(\{C_{2}|| \mathcal{S}_3\}\) symmetries are broken and the spin degeneracy is lifted, as shown in Figs. \ref{fig:4}(c) and  \ref{fig:4}(d). The band exhibits ferromagnetic splitting while the overall magnetic moment of the system remains zero, a phenomenon known as fully compensated ferrimagnetism~\cite{liuTwoDimensionalFullyCompensated2025a}. The \(\mathcal{E}_z\)-induced ferromagnetic band structure and ferromagnetic properties characterize monolayer Sr\(_3\)Fe\(_2\)O\(_5\)Cl\(_2\) as Class-III.

As an effective probe of magnetic moments, MOKE can also reveal the properties of Class-III~\cite{ebertMagnetoopticalEffectsTransition1996,guoBandtheoreticalInvestigationMagnetooptical1995,OPPENEER2001229,sangalliPseudopotentialbasedFirstprinciplesApproach2012}. When SOC is included, the \(\mathcal{PT}\) symmetry of Sr$_3$Fe$_2$O$_5$Cl$_2$ simultaneously protects the band spin degeneracy and forbids all Hall photoconductivity components; consequently, the pristine system exhibits no MOKE (see details in SM\cite{SM}). Since monolayer Sr$_3$Fe$_2$O$_5$Cl$_2$ is a soft magnet (see details in SM\cite{SM}), its magnetization direction can be easily manipulated. When its magnetic moment is oriented along [001] and an \(\mathcal{E}_z\) is applied, the \(\mathcal{PT}\) symmetry is broken, pronounced spin splitting appears [Fig.~\ref{fig:4}(c)], and MOKE emerges. The calculated results for \(\mathcal{E}_z=0.2\,\mathrm{eV}/\text{\AA}\) are shown in Fig.~\ref{fig:4}(e); when the electric field is reversed to \(\mathcal{E}_z=-0.2\,\mathrm{eV}/\text{\AA}\), the band exhibits opposite chiral splitting and MOKE undergoes a sign reversal as well [Figs.~\ref{fig:4}(d) and (f)].

\textit{\textcolor{blue}{Conclusion.}}---We have unveiled a hidden landscape of unconventional collinear magnetism in 2D \(\mathcal{PT}\)-symmetric antiferromagnets, where the combined \(\mathcal{PT}\) symmetry has long obscured the role of other crystallographic symmetries in shaping the electronic structure. By constructing the full classification of 88 cSLGs, we demonstrate that an \(\mathcal{E}_z\) serves as a universal knob to release these hidden constraints, yielding three exhaustive classes of emergent magnetic phases. Notably, the distinct spin-splitting patterns and spin-dependent transport responses, including electric Hall effect, spin-dependent transport, and electric reversible MOKE, are not accidental but are dictated by the underlying symmetry of each class. Our work thus establishes a symmetry-based taxonomy for electric hidden unconventional magnetism, transforming what was previously a set of disparate phenomena into a coherent, complete, and predictive framework. The identified candidate materials, all amenable to exfoliation, provide an experimentally accessible testbed for realizing and controlling these phases; moreover, our classification scheme can be readily extended to other layered magnets with similar hidden-symmetry structures.


\bibliography{ref}

@article{kurebayashiMagnetismSymmetrySpin2022,
	title = {{M}agnetism, {S}ymmetry and {S}pin {T}ransport in van {D}er {{Waals}} {L}ayered {S}ystems},
	author = {Kurebayashi, Hidekazu and Garcia, Jose H. and Khan, Safe and Sinova, Jairo and Roche, Stephan},
	year = 2022,
	month = mar,
	journal = {Nat. Rev. Phys.},
	volume = {4},
	number = {3},
	pages = {150--166},
	issn = {2522-5820}
}

@article{yuanGiantMomentumdependentSpin2020b,
	title = {{G}iant {M}omentum-{D}ependent {S}pin {S}plitting in {C}entrosymmetric {L}ow-{Z}     {A}ntiferromagnets},
	author = {Yuan, Lin-Ding and Wang, Zhi and Luo, Jun-Wei and Rashba, Emmanuel I. and Zunger, Alex},
	year = 2020,
	month = jul,
	journal = {Phys. Rev. B},
	volume = {102},
	number = {1},
	pages = {014422},
	issn = {2469-9950, 2469-9969},
	urldate = {2026-09-28},
	langid = {english}
}

@article{yuanPredictionLowZCollinear2021,
	title = {{P}rediction of {L}ow-{{Z}} {C}ollinear and {N}oncollinear {A}ntiferromagnetic {C}ompounds {H}aving {M}omentum-{D}ependent {S}pin {S}plitting {E}ven without {S}pin-{O}rbit {C}oupling},
	author = {Yuan, Lin-Ding and Wang, Zhi and Luo, Jun-Wei and Zunger, Alex},
	year = 2021,
	month = jan,
	journal = {Phys. Rev. Mater.},
	volume = {5},
	number = {1},
	pages = {014409},
	issn = {2475-9953},
	urldate = {2026-09-28},
	langid = {english}
}

@article{mazinEditorialAltermagnetismNew2022a,
	title = {{E}ditorial: {{Altermagnetism}}---{{A New Punch Line}} of {{Fundamental Magnetism}}},
	shorttitle = {Editorial},
	author = {Mazin, Igor and {The PRX Editors}},
	year = 2022,
	month = dec,
	journal = {Phys. Rev. X},
	volume = {12},
	number = {4},
	pages = {040002},
	issn = {2160-3308},
	urldate = {2026-09-28},
	langid = {english}
}

@article{DEGROOT199145,
	title = {{H}alf-{M}etallic {M}agnetism in the 1990s},
	author = {{de Groot}, Robert A.},
	year = 1991,
	journal = {Physica B},
	volume = {172},
	number = {1},
	pages = {45--50},
	issn = {0921-4526}
}

@article{degrootNewClassMaterials1983,
	title = {{N}ew {{Class}} of {{Materials}}: {{Half-Metallic Ferromagnets}}},
	shorttitle = {New {{Class}} of {{Materials}}},
	author = {De Groot, R. A. and Mueller, F. M. and Engen, P. G. Van and Buschow, K. H. J.},
	year = 1983,
	month = jun,
	journal = {Phys. Rev. Lett.},
	volume = {50},
	number = {25},
	pages = {2024--2027},
	issn = {0031-9007},
	urldate = {2026-09-28},
	copyright = {http://link.aps.org/licenses/aps-default-license},
	langid = {english}
}

@article{huHalfMetallicAntiferromagnetProspective2012,
	title = {{H}alf-{{Metallic Antiferromagnet}} as a {{Prospective Material}} for {{Spintronics}}},
	author = {Hu, Xiao},
	year = 2012,
	journal = {Adv. Mater.},
	volume = {24},
	number = {2},
	pages = {294--298},
	issn = {1521-4095},
	urldate = {2026-09-28},
	langid = {english}
}

@article{katsnelsonHalfmetallicFerromagnetsBand2008a,
	title = {{H}alf-{M}etallic {F}erromagnets: {{From}} {B}and {S}tructure to {M}any-{B}ody {E}ffects},
	shorttitle = {Half-Metallic Ferromagnets},
	author = {Katsnelson, M. I. and Irkhin, V. {\relax Yu}. and Chioncel, L. and Lichtenstein, A. I. and De Groot, R. A.},
	year = 2008,
	month = apr,
	journal = {Rev. Mod. Phys.},
	volume = {80},
	number = {2},
	pages = {315--378},
	issn = {0034-6861, 1539-0756},
	urldate = {2026-09-28},
	copyright = {http://link.aps.org/licenses/aps-default-license},
	langid = {english}
}

@article{rimmlerNoncollinearAntiferromagneticSpintronics2025,
	title = {{N}on-{C}ollinear {A}ntiferromagnetic {S}pintronics},
	author = {Rimmler, Berthold H. and Pal, Banabir and Parkin, Stuart S. P.},
	year = 2025,
	month = feb,
	journal = {Nat. Rev. Mater.},
	volume = {10},
	number = {2},
	pages = {109--127},
	issn = {2058-8437}
}

@article{vanleukenHalfMetallicAntiferromagnets1995a,
	title = {{H}alf-{{Metallic Antiferromagnets}}},
	author = {Van Leuken, H. and De Groot, R. A.},
	year = 1995,
	month = feb,
	journal = {Phys. Rev. Lett.},
	volume = {74},
	number = {7},
	pages = {1171--1173},
	issn = {0031-9007, 1079-7114},
	urldate = {2026-09-28},
	copyright = {http://link.aps.org/licenses/aps-default-license},
	langid = {english}
}

@article{luoSpinGroupSymmetry2026,
	title = {{S}pin {{Group Symmetry Criteria}} for {{Odd-parity Magnets}}},
	author = {Luo, Xun-Jiang and Hu, Jin-Xin and Hu, Meng-Li and Law, K. T.},
	year = 2026,
	month = jan,
	journal = {arXiv:2510.05512},

}

@article{hellenes2023p,
	title={{P}-wave magnets},
	author={Hellenes, Anna Birk and Jungwirth, Tom{\'a}{\v{s}} and Jaeschke-Ubiergo, Rodrigo and Chakraborty, Atasi and Sinova, Jairo and {\v{S}}mejkal, Libor},
	journal={arXiv:2309.01607},
	year={2023}
}

@article{brekke2024minimal,
	title={{M}inimal models and transport properties of unconventional p-wave magnets},
	author={Brekke, Bj{\o}rnulf and Sukhachov, Pavlo and Giil, Hans Gl{\o}ckner and Brataas, Arne and Linder, Jacob},
	journal={Phys. Rev. Lett.},
	volume={133},
	number={23},
	pages={236703},
	year={2024},
	publisher={APS}
}

@article{PhysRevLettweng,
	title = {Magnetization-Induced Band Shift in Ferromagnetic Weyl Semimetal {{Co{$_3$}Sn{$_2$}S{$_2$}}}},
	author = {Yang, Run and Zhang, Tan and Zhou, Liqin and Dai, Yaomin and Liao, Zhiyu and Weng, Hongming and Qiu, Xianggang},
	journal = {Phys. Rev. Lett.},
	volume = {124},
	issue = {7},
	pages = {077403},
	numpages = {6},
	year = {2020},
	month = {Feb},
}

@article{liStackingdependentElectronicTopological2025,
	title = {Stacking-Dependent Electronic and Topological Properties in van Der {{Waals}} Antiferromagnet {{MnBi{$_2$}Te{$_4$}}} Films},
	author = {Li, Jiaheng and Wu, Quansheng and Weng, Hongming},
	year = 2025,
	month = mar,
	journal = {npj Comput. Mater.},
	volume = {11},
	number = {1},
	pages = {53},
	issn = {2057-3960}
}

@article{maMultifunctionalAntiferromagneticMaterials2021c,
	title = {{M}ultifunctional {A}ntiferromagnetic {M}aterials with {G}iant {P}iezomagnetism and {N}oncollinear {S}pin {C}urrent},
	author = {Ma, Hai-Yang and Hu, Mengli and Li, Nana and Liu, Jianpeng and Yao, Wang and Jia, Jin-Feng and Liu, Junwei},
	year = 2021,
	month = may,
	journal = {Nat. Commun.},
	volume = {12},
	number = {1},
	pages = {2846},
	issn = {2041-1723}
}

@article{zhouManipulationAltermagneticOrder2025,
	title = {{M}anipulation of the {A}ltermagnetic {O}rder in {{CrSb}} via {C}rystal {S}ymmetry},
	author = {Zhou, Zhiyuan and Cheng, Xingkai and Hu, Mengli and Chu, Ruiyue and Bai, Hua and Han, Lei and Liu, Junwei and Pan, Feng and Song, Cheng},
	year = 2025,
	month = feb,
	journal = {Nature},
	volume = {638},
	number = {8051},
	pages = {645--650},
	issn = {1476-4687}
}

@article{sciadv.adn0479,
	title = {{E}lectrical 180{$^\circ$} {S}witching of {{N\'eel}} {V}ector in {S}pin-{S}plitting {A}ntiferromagnet},
	author = {Han, Lei and Fu, Xizhi and Peng, Rui and Cheng, Xingkai and Dai, Jiankun and Liu, Liangyang and Li, Yidian and Zhang, Yichi and Zhu, Wenxuan and Bai, Hua and Zhou, Yongjian and Liang, Shixuan and Chen, Chong and Wang, Qian and Chen, Xianzhe and Yang, Luyi and Zhang, Yang and Song, Cheng and Liu, Junwei and Pan, Feng},
	year = 2024,
	journal = {Sci. Adv.},
	volume = {10},
	number = {4},
	pages = {eadn0479}
}

@article{yuQuantizedAnomalousHall2010a,
	title = {{Q}uantized {{Anomalous Hall Effect}} in {{Magnetic Topological Insulators}}},
	author = {Yu, Rui and Zhang, Wei and Zhang, Hai-Jun and Zhang, Shou-Cheng and Dai, Xi and Fang, Zhong},
	year = 2010,
	month = jul,
	journal = {Science},
	volume = {329},
	number = {5987},
	pages = {61--64},
	issn = {0036-8075, 1095-9203},
	urldate = {2026-09-28},
	langid = {english}
}

@article{chenIntrinsicMagneticTopological2019,
	title = {{I}ntrinsic {M}agnetic {T}opological {I}nsulator {P}hases in the {{Sb}} {D}oped {{MnBi2Te4}} {B}ulks and {T}hin {F}lakes},
	author = {Chen, Bo and Fei, Fucong and Zhang, Dongqin and Zhang, Bo and Liu, Wanling and Zhang, Shuai and Wang, Pengdong and Wei, Boyuan and Zhang, Yong and Zuo, Zewen and Guo, Jingwen and Liu, Qianqian and Wang, Zilu and Wu, Xuchuan and Zong, Junyu and Xie, Xuedong and Chen, Wang and Sun, Zhe and Wang, Shancai and Zhang, Yi and Zhang, Minhao and Wang, Xuefeng and Song, Fengqi and Zhang, Haijun and Shen, Dawei and Wang, Baigeng},
	year = 2019,
	month = oct,
	journal = {Nat. Commun.},
	volume = {10},
	number = {1},
	pages = {4469},
	publisher = {Nature Publishing Group},
	issn = {2041-1723},
	urldate = {2026-09-28},
	copyright = {2019 The Author(s)},
	langid = {english}
}

@article{zhanFloquetEngineeringNonequilibrium2023,
	title = {{F}loquet {{Engineering}} of {{Nonequilibrium Valley-Polarized Quantum Anomalous Hall Effect}} with {{Tunable Chern Number}}},
	author = {Zhan, Fangyang and Zeng, Junjie and Chen, Zhuo and Jin, Xin and Fan, Jing and Chen, Tingyong and Wang, Rui},
	year = 2023,
	month = mar,
	journal = {Nano Lett.},
	volume = {23},
	number = {6},
	pages = {2166--2172},
	issn = {1530-6984, 1530-6992},
	urldate = {2026-09-28},
	copyright = {https://doi.org/10.15223/policy-029},
	langid = {english}
}

@article{zhanPerspectiveFloquetEngineering2024,
	title = {{P}erspective: {{Floquet}} {E}ngineering {T}opological {S}tates from {E}ffective {M}odels towards {R}ealistic {M}aterials},
	shorttitle = {Perspective},
	author = {Zhan, Fangyang and Chen, Rui and Ning, Zhen and Ma, Da-Shuai and Wang, Ziming and Xu, Dong-Hui and Wang, Rui},
	year = 2024,
	month = oct,
	journal = {Quantum Front.},
	volume = {3},
	number = {1},
	pages = {21},
	issn = {2731-6106},
	urldate = {2026-09-28},
	langid = {english}
}

@article{smejkalConventionalFerromagnetismAntiferromagnetism2022f,
	title = {{B}eyond {{Conventional Ferromagnetism}} and {{Antiferromagnetism}}: {{A Phase}} with {{Nonrelativistic Spin}} and {{Crystal Rotation Symmetry}}},
	shorttitle = {Beyond {{Conventional Ferromagnetism}} and {{Antiferromagnetism}}},
	author = {{\v S}mejkal, Libor and Sinova, Jairo and Jungwirth, Tomas},
	year = 2022,
	month = sep,
	journal = {Phys. Rev. X},
	volume = {12},
	number = {3},
	pages = {031042},
	publisher = {American Physical Society},
	urldate = {2026-03-12}
}

@article{smejkalEmergingResearchLandscape2022b,
	title = {{E}merging {{Research Landscape}} of {{Altermagnetism}}},
	author = {{\v S}mejkal, Libor and Sinova, Jairo and Jungwirth, Tomas},
	year = 2022,
	month = dec,
	journal = {Phys. Rev. X},
	volume = {12},
	number = {4},
	pages = {040501},
	issn = {2160-3308},
	urldate = {2026-09-28},
	langid = {english}
}

@article{duanAntiferroelectricAltermagnetsAntiferroelectricity2025a,
	title = {{A}ntiferroelectric {{Altermagnets}}: {{Antiferroelectricity Alters Magnets}}},
	shorttitle = {Antiferroelectric {{Altermagnets}}},
	author = {Duan, Xunkai and Zhang, Jiayong and Zhu, Ziye and Liu, Yuntian and Zhang, Zhenyu and {\v Z}uti{\'c}, Igor and Zhou, Tong},
	year = 2025,
	month = mar,
	journal = {Phys. Rev. Lett.},
	volume = {134},
	number = {10},
	pages = {106801},
	issn = {0031-9007, 1079-7114},
	urldate = {2026-09-28},
	langid = {english}
}

@article{zhuAltermagneticProximityEffect2026,
	title = {{A}ltermagnetic {{Proximity Effect}}},
	author = {Zhu, Ziye and Huang, Richang and Chen, Xianzhang and Cui, Zhou and Duan, Xunkai and Zhang, Jiayong and {\v Z}uti{\'c}, Igor and Zhou, Tong},
	year = 2026,
	month = may,
	journal = {Phys. Rev. Lett.},
	volume = {136},
	number = {18},
	pages = {186702},
	publisher = {American Physical Society},
	urldate = {2026-09-28}
}

@article{shaoNeelSpinCurrents2023a,
	title = {{N\'eel} {{Spin Currents}} in {{Antiferromagnets}}},
	author = {Shao, Ding-Fu and Jiang, Yuan-Yuan and Ding, Jun and Zhang, Shu-Hui and Wang, Zi-An and Xiao, Rui-Chun and Gurung, Gautam and Lu, W. J. and Sun, Y. P. and Tsymbal, Evgeny Y.},
	year = 2023,
	month = may,
	journal = {Phys. Rev. Lett.},
	volume = {130},
	number = {21},
	pages = {216702},
	publisher = {American Physical Society},
	urldate = {2026-09-28}
}

@article{zhangXtypeStackingCrosschain2025,
	title = {{X}-{T}ype {S}tacking in {C}ross-{C}hain {A}ntiferromagnets},
	author = {Zhang, Shui-Sen and Wang, Zi-An and Li, Bo and Jiang, Yuan-Yuan and Zhang, Shu-Hui and Xiao, Rui-Chun and Liu, Lan-Xin and Luo, Xuan and Lu, Wen-Jian and Tian, Mingliang and Sun, Yu-Ping and Tsymbal, Evgeny Y. and Du, Haifeng and Shao, Ding-Fu},
	year = 2025,
	month = jun,
	journal = {Newton},
	volume = {1},
	number = {4},
	pages = {100068},
	issn = {29506360},
	urldate = {2026-09-28},
	langid = {english}
}

@article{fengDiscoveryWeylNodal2019a,
	title = {{D}iscovery of {{Weyl Nodal Lines}} in a {{Single-Layer Ferromagnet}}},
	author = {Feng, Baojie and Zhang, Run-Wu and Feng, Ya and Fu, Botao and Wu, Shilong and Miyamoto, Koji and He, Shaolong and Chen, Lan and Wu, Kehui and Shimada, Kenya and Okuda, Taichi and Yao, Yugui},
	year = 2019,
	month = sep,
	journal = {Phys. Rev. Lett.},
	volume = {123},
	number = {11},
	pages = {116401},
	issn = {0031-9007, 1079-7114},
	urldate = {2026-09-23},
	langid = {english}
}

@article{zhangNodalLineSpinGapless2020b,
	title = {{N}odal {{Line Spin-Gapless Semimetals}} and {{High-Quality Candidate Materials}}},
	author = {Zhang, Run-Wu and Zhang, Zeying and Liu, Cheng-Cheng and Yao, Yugui},
	year = 2020,
	month = jan,
	journal = {Phys. Rev. Lett.},
	volume = {124},
	number = {1},
	pages = {016402},
	publisher = {American Physical Society},
	urldate = {2026-09-23}
}

@article{SM,
	journal = {See Supplemental Material for details about spin layer groups, candidate materials and DFT calculations},
}

@article{alexanderSymmetryConsiderationsDetermination1962,
	title = {{S}ymmetry {{Considerations}} in the {{Determination}} of {{Magnetic Structures}}},
	author = {Alexander, S.},
	year = 1962,
	month = jul,
	journal = {Phys. Rev.},
	volume = {127},
	number = {2},
	pages = {420--432},
	issn = {0031-899X},
	urldate = {2026-09-23},
	copyright = {http://link.aps.org/licenses/aps-default-license},
	langid = {english}
}

@article{dimmockUseSymmetryDetermination1963,
	title = {{U}se of {{Symmetry}} in the {{Determination}} of {{Magnetic Structures}}},
	author = {Dimmock, John O.},
	year = 1963,
	month = may,
	journal = {Phys. Rev.},
	volume = {130},
	number = {4},
	pages = {1337--1344},
	publisher = {American Physical Society},
	urldate = {2026-09-23}
}

@article{donnaySymmetryMagneticStructures1958,
	title = {{S}ymmetry of {{Magnetic Structures}}: {{Magnetic Structure}} of {{Chalcopyrite}}},
	shorttitle = {Symmetry of {{Magnetic Structures}}},
	author = {Donnay, G. and Corliss, L. M. and Donnay, J. D. H. and Elliott, N. and Hastings, J. M.},
	year = 1958,
	month = dec,
	journal = {Phys. Rev.},
	volume = {112},
	number = {6},
	pages = {1917--1923},
	issn = {0031-899X},
	urldate = {2026-09-23},
	copyright = {http://link.aps.org/licenses/aps-default-license},
	langid = {english}
}

@article{tokuraMagneticTopologicalInsulators2019,
	title = {{M}agnetic {T}opological {I}nsulators},
	author = {Tokura, Yoshinori and Yasuda, Kenji and Tsukazaki, Atsushi},
	year = 2019,
	month = feb,
	journal = {Nat. Rev. Phys.},
	volume = {1},
	number = {2},
	pages = {126--143},
	issn = {2522-5820}
}

@article{deng2020quantum,
	title={{Q}uantum anomalous {H}all effect in intrinsic magnetic topological insulator {MnBi2Te4}},
	author={Deng, Yujun and Yu, Yijun and Shi, Meng Zhu and Guo, Zhongxun and Xu, Zihan and Wang, Jing and Chen, Xian Hui and Zhang, Yuanbo},
	journal={Science},
	volume={367},
	number={6480},
	pages={895--900},
	year={2020},
	publisher={American Association for the Advancement of Science}
}

@article{elcoroMagneticTopologicalQuantum2021,
	title = {{M}agnetic {T}opological {Q}uantum {C}hemistry},
	author = {Elcoro, Luis and Wieder, Benjamin J. and Song, Zhida and Xu, Yuanfeng and Bradlyn, Barry and Bernevig, B. Andrei},
	year = 2021,
	month = oct,
	journal = {Nat. Commun.},
	volume = {12},
	number = {1},
	pages = {5965},
	publisher = {Nature Publishing Group},
	issn = {2041-1723},
	urldate = {2026-09-27},
	copyright = {2021 The Author(s)},
	langid = {english}
}

@article{bernevigProgressProspectsMagnetic2022,
	title = {{P}rogress and {P}rospects in {M}agnetic {T}opological {M}aterials},
	author = {Bernevig, B. Andrei and Felser, Claudia and Beidenkopf, Haim},
	year = 2022,
	month = mar,
	journal = {Nature},
	volume = {603},
	number = {7899},
	pages = {41--51},
	publisher = {Nature Publishing Group},
	issn = {1476-4687},
	urldate = {2026-09-27},
	copyright = {2022 The Author(s), under exclusive licence to Springer Nature Limited},
	langid = {english}
}

@article{zhangMagneticTopologicalMaterials2023,
	title = {{M}agnetic {T}opological {M}aterials in {T}wo-{D}imensional: {T}heory, {M}aterial {R}ealization and {A}pplication {P}rospects},
	shorttitle = {Magnetic Topological Materials in Two-Dimensional},
	author = {Zhang, Xiaoming and Wang, Xiaotian and He, Tingli and Wang, Lirong and Yu, Wei-Wang and Liu, Ying and Liu, Guodong and Cheng, Zhenxiang},
	year = 2023,
	month = nov,
	journal = {Sci. Bull.},
	volume = {68},
	number = {21},
	pages = {2639--2657},
	issn = {2095-9273},
	urldate = {2026-09-27}
}

@article{wuNodalSurfaceSemimetals2018a,
	title = {{N}odal {S}urface {S}emimetals: {{Theory}} and {M}aterial {R}ealization},
	shorttitle = {Nodal Surface Semimetals},
	author = {Wu, Weikang},
	year = 2018,
	journal = {Phys. Rev. B},
	volume = {97},
    pages={115125},
	number = {11}
}

@article{PhysRevLett.128.066602,
	title = {{B}icircular {L}ight {F}loquet {E}ngineering of {M}agnetic {S}ymmetry and {T}opology and {I}ts {A}pplication to the {D}irac {S}emimetal {{Cd}}{$_{3}$}{{As}}{$_{2}$}},
	author = {Trevisan, Tha{\'\i}s V. and Arribi, Pablo Villar and Heinonen, Olle and Slager, Robert-Jan and Orth, Peter P.},
	year = 2022,
	month = feb,
	journal = {Phys. Rev. Lett.},
	volume = {128},
	number = {6},
	pages = {066602},
	publisher = {American Physical Society}
}

@book{bradley1972,
	author    = {Bradley, C. J. and Cracknell, A. P.},
	title     = {{T}he {M}athematical {T}heory of {S}ymmetry in {S}olids: {R}epresentation {T}heory for {P}oint {G}roups and {S}pace {G}roups},
	publisher = {Oxford University Press},
	address   = {New York},
	year      = {1972}
}

@article{chenEnumerationRepresentationTheory2024f,
	title = {{E}numeration and {{Representation Theory}} of {{Spin Space Groups}}},
	author = {Chen, Xiaobing and Ren, Jun and Zhu, Yanzhou and Yu, Yutong and Zhang, Ao and Liu, Pengfei and Li, Jiayu and Liu, Yuntian and Li, Caiheng and Liu, Qihang},
	year = 2024,
	month = aug,
	journal = {Phys. Rev. X},
	volume = {14},
	number = {3},
	pages = {031038},
	publisher = {American Physical Society},
	urldate = {2026-03-12}
}

@article{jiangEnumerationSpinSpaceGroups2024c,
	title = {{E}numeration of {{Spin-Space Groups}}: {{Toward}} a {{Complete Description}} of {{Symmetries}} of {{Magnetic Orders}}},
	shorttitle = {Enumeration of {{Spin-Space Groups}}},
	author = {Jiang, Yi and Song, Ziyin and Zhu, Tiannian and Fang, Zhong and Weng, Hongming and Liu, Zheng-Xin and Yang, Jian and Fang, Chen},
	year = 2024,
	month = aug,
	journal = {Phys. Rev. X},
	volume = {14},
	number = {3},
	pages = {031039},
	publisher = {American Physical Society},
	urldate = {2026-09-23}
}

@article{xiaoSpinSpaceGroups2024b,
	title = {{S}pin {{Space Groups}}: {{Full Classification}} and {{Applications}}},
	shorttitle = {Spin {{Space Groups}}},
	author = {Xiao, Zhenyu and Zhao, Jianzhou and Li, Yanqi and Shindou, Ryuichi and Song, Zhi-Da},
	year = 2024,
	month = aug,
	journal = {Phys. Rev. X},
	volume = {14},
	number = {3},
	pages = {031037},
	publisher = {American Physical Society},
	urldate = {2026-09-23}
}

@article{liuSymmetryClassificationMagnetic2026,
	title = {{S}ymmetry {C}lassification of {M}agnetic {O}rders {U}sing {O}riented {S}pin {S}pace {G}roups},
	author = {Liu, Yuntian and Chen, Xiaobing and Yu, Yutong and Etxebarria, Jes{\'u}s and {Perez-Mato}, J. Manuel and Liu, Qihang},
	year = 2026,
	month = apr,
	journal = {Nature},
	volume = {652},
	number = {8111},
	pages = {869--873},
	issn = {0028-0836, 1476-4687},
	urldate = {2026-09-23},
	langid = {english}
}

@article{chengSpinPumpingSpinTransfer2014a,
	title = {{S}pin {{Pumping}} and {{Spin-Transfer Torques}} in {{Antiferromagnets}}},
	author = {Cheng, Ran and Xiao, Jiang and Niu, Qian and Brataas, Arne},
	year = 2014,
	month = jul,
	journal = {Phys. Rev. Lett.},
	volume = {113},
	number = {5},
	pages = {057601},
	issn = {0031-9007, 1079-7114},
	urldate = {2026-09-24},
	copyright = {http://link.aps.org/licenses/aps-default-license},
	langid = {english}
}

@article{danielsSpintransferTorqueInduced2015,
	title = {{S}pin-{T}ransfer {T}orque {I}nduced {S}pin {W}aves in {A}ntiferromagnetic {I}nsulators},
	author = {Daniels, Matthew W and Guo, Wei and Stocks, G Malcolm and Xiao, Di and Xiao, Jiang},
	year = 2015,
	month = oct,
	journal = {New J. Phys.},
	volume = {17},
	number = {10},
	pages = {103039},
	issn = {1367-2630},
	urldate = {2026-09-24}
}

@article{jungwirthAntiferromagneticSpintronics2016b,
	title = {{A}ntiferromagnetic {S}pintronics},
	author = {Jungwirth, T. and Marti, X. and Wadley, P. and Wunderlich, J.},
	year = 2016,
	month = mar,
	journal = {Nat. Nanotechnol.},
	volume = {11},
	number = {3},
	pages = {231--241},
	issn = {1748-3387, 1748-3395},
	urldate = {2026-09-24},
	langid = {english}
}

@incollection{Šmejkal2018,
	title = {{S}ymmetry and {T}opology in {A}ntiferromagnetic {S}pintronics},
	booktitle = {Topology in Magnetism},
	author = {{\v S}mejkal, Libor and Jungwirth, Tom{\'a}{\v s}},
	editor = {Zang, Jiadong and Cros, Vincent and Hoffmann, Axel},
	year = 2018,
	pages = {267--298},
	publisher = {Springer International Publishing},
	address = {Cham},
	isbn = {978-3-319-97334-0}
}

@article{tangDiracFermionsAntiferromagnetic2016,
	title = {{D}irac {F}ermions in an {A}ntiferromagnetic {S}emimetal},
	author = {Tang, Peizhe and Zhou, Quan and Xu, Gang and Zhang, Shou-Cheng},
	year = 2016,
	month = dec,
	journal = {Nat. Phys.},
	volume = {12},
	number = {12},
	pages = {1100--1104},
	issn = {1745-2473, 1745-2481},
	urldate = {2025-09-18},
	langid = {english}
}

@article{watanabeMagneticParityViolation2024,
	title = {{M}agnetic {P}arity {V}iolation and {P}arity-{T}ime-{R}eversal-{S}ymmetric {M}agnets},
	author = {Watanabe, Hikaru and Yanase, Youichi},
	year = 2024,
	month = sep,
	journal = {J. Phys.: Condens. Matter},
	volume = {36},
	number = {37},
	pages = {373001},
	issn = {0953-8984, 1361-648X},
	urldate = {2026-09-23}
}

@article{zhangParityTimeSymmetryMagnetic2024,
	title = {{P}arity-{{Time Symmetry}} in {{Magnetic Materials}} and {{Devices}}},
	author = {Zhang, Zhitao and Xin, Chao and Liu, Haoliang},
	year = 2024,
	journal = {Adv. Electron. Mater.},
	volume = {10},
	number = {3},
	pages = {2300674},
	issn = {2199-160X},
	urldate = {2026-09-23},
	langid = {english}
}

@article{sun2025multifield,
	title={{M}ultifield-induced antiferromagnet transformation into altermagnet and realized anomalous valley {H}all effect in monolayer {VPSe}$_3$},
	author={Sun, Hanbo and Dong, Pengqiang and Wu, Chao and Li, Ping},
	journal={Phys. Rev. B},
	volume={111},
	number={23},
	pages={235431},
	year={2025},
	publisher={APS}
}

@article{du2020berry,
	title={{B}erry curvature engineering by gating two-dimensional antiferromagnets},
	author={Du, Shiqiao and Tang, Peizhe and Li, Jiaheng and Lin, Zuzhang and Xu, Yong and Duan, Wenhui and Rubio, Angel},
	journal={Phys. Rev. Res.},
	volume={2},
	number={2},
	pages={022025},
	year={2020},
	publisher={APS}
}

@article{z69z-175r,
	title = {{A}nomalous {H}all effect driven by electric field and sliding ferroelectricity in two-dimensional compensated antiferromagnetic {MnS}},
	author = {Long, Guozhi and Huang, Huaqing},
	journal = {Phys. Rev. B},
	volume = {112},
	issue = {16},
	pages = {165136},
	numpages = {7},
	year = {2025},
	month = {Oct},
	publisher = {American Physical Society},
}

@article{qayyum2024pt,
	title={{P}t-symmetry breaking and spin control in {2D} antiferromagnetic {MnSe}},
	author={Qayyum, Hafiz Adil and Mansha, Muhammad and Sattar, Shahid},
	journal={ACS Omega},
	volume={9},
	number={47},
	pages={47097},
	year={2024}
}

@article{guo2024electric,
	title={{E}lectric-field-tuned anomalous valley {H}all effect in {A}-type hexagonal antiferromagnetic monolayers},
	author={Guo, San-Dong and Tao, Yu-Ling and Zhuo, Zi-Yang and Zhu, Gangqiang and Ang, Yee Sin},
	journal={Phys. Rev. B},
	volume={109},
	number={13},
	pages={134402},
	year={2024},
	publisher={APS}
}

@article{wang2025electric,
	title={{E}lectric-field-induced switchable two-dimensional altermagnets},
	author={Wang, Dinghui and Wang, Huaiqiang and Liu, Lulu and Zhang, Junting and Zhang, Haijun},
	journal={Nano Lett.},
	volume={25},
	number={1},
	pages={498--503},
	year={2025},
	publisher={ACS Publications}
}

@article{zhangSpinLayerGroups2026,
	title = {{S}pin {L}ayer {G}roups and {T}heir {C}orepresentations},
	author = {Zhang, Zeying and Liu, Gui-Bin and Tian, Mu and Zhang, Run-Wu and Yu, Zhi-Ming and Yao, Yugui},
	year = 2026,
	month = may,
	journal = {arXiv:2605.25484}
}

@article{baiAnomalousHallEffect2025f,
	title = {{A}nomalous {{Hall Effect}} in {{Type IV 2D Collinear Magnets}}},
	author = {Bai, Ling and Zhang, Run-Wu and Feng, Wanxiang and Yao, Yugui},
	year = 2025,
	month = jul,
	journal = {Phys. Rev. Lett.},
	volume = {135},
	number = {3},
	pages = {036702},
	issn = {0031-9007, 1079-7114},
	urldate = {2026-09-23},
	langid = {english}
}

@article{tianSymmetryClassificationAltermagnetism2026,
	title = {{S}ymmetry {{Classification}} of {{Altermagnetism}} and {{Emergence}} of {{Type-IV Magnetism}} in {{Two Dimensions}}},
	author = {Tian, Mu and Cui, Chaoxi and Zhang, Zeying and Duan, Jingyi and Feng, Wanxiang and Zhang, Run-Wu},
	year = 2026,
	month = may,
	journal = {Phys. Rev. Lett.},
	volume = {136},
	number = {20},
	pages = {206701},
	issn = {0031-9007, 1079-7114},
	urldate = {2026-07-23},
	langid = {english}
}

@article{chenUnconventionalMagnonsCollinear2025b,
	title = {{U}nconventional {M}agnons in {C}ollinear {M}agnets {D}ictated by {S}pin {S}pace {G}roups},
	author = {Chen, Xiaobing and Liu, Yuntian and Liu, Pengfei and Yu, Yutong and Ren, Jun and Li, Jiayu and Zhang, Ao and Liu, Qihang},
	year = 2025,
	month = apr,
	journal = {Nature},
	volume = {640},
	number = {8058},
	pages = {349--354},
	issn = {0028-0836, 1476-4687},
	urldate = {2025-09-02},
	langid = {english}
}

@article{cuiElectricHallEffect2025,
	title = {{E}lectric {{Hall Effect}} and {{Quantum Electric Hall Effect}}},
	author = {Cui, Chaoxi and Zhang, Run-Wu and Qiu, Yuhui and Han, Yilin and Yu, Zhi-Ming and Yao, Yugui},
	year = 2025,
	month = sep,
	journal = {Phys. Rev. Lett.},
	volume = {135},
	number = {11},
	pages = {116301},
	issn = {0031-9007, 1079-7114},
	urldate = {2025-12-15},
	langid = {english}
}

@article{litvinSpinGroups1974,
	title = {{S}pin {G}roups},
	author = {Litvin, D. B. and Opechowski, W.},
	year = 1974,
	month = sep,
	journal = {Physica},
	volume = {76},
	number = {3},
	pages = {538--554},
	issn = {0031-8914},
	urldate = {2026-09-24}
}

@article{litvinSpinPointGroups1977a,
	title = {{S}pin {P}oint {G}roups},
	author = {Litvin, D. B.},
	year = 1977,
	month = mar,
	journal = {Acta Crystallogr. Sect. A},
	volume = {33},
	number = {2},
	pages = {279--287},
	issn = {0567-7394},
	urldate = {2026-09-24},
	copyright = {http://journals.iucr.org/services/copyrightpolicy.html},
	langid = {english}
}

@article{liuTwoDimensionalFullyCompensated2025a,
	title = {{T}wo-{{Dimensional Fully Compensated Ferrimagnetism}}},
	author = {Liu, Yichen and Guo, San-Dong and Li, Yongpan and Liu, Cheng-Cheng},
	year = 2025,
	month = mar,
	journal = {Phys. Rev. Lett.},
	volume = {134},
	number = {11},
	pages = {116703},
	publisher = {American Physical Society},
	urldate = {2026-04-09}
}

@article{matsukuraControlMagnetismElectric2015,
	title = {{C}ontrol of {M}agnetism by {E}lectric {F}ields},
	author = {Matsukura, Fumihiro and Tokura, Yoshinori and Ohno, Hideo},
	year = 2015,
	month = mar,
	journal = {Nat. Nanotechnol.},
	volume = {10},
	number = {3},
	pages = {209--220},
	issn = {1748-3395}
}

@article{kajiwaraTransmissionElectricalSignals2010,
	title = {{T}ransmission of {E}lectrical {S}ignals by {S}pin-{W}ave {I}nterconversion in a {M}agnetic {I}nsulator},
	author = {Kajiwara, Y. and Harii, K. and Takahashi, S. and Ohe, J. and Uchida, K. and Mizuguchi, M. and Umezawa, H. and Kawai, H. and Ando, K. and Takanashi, K. and Maekawa, S. and Saitoh, E.},
	year = 2010,
	month = mar,
	journal = {Nature},
	volume = {464},
	number = {7286},
	pages = {262--266},
	issn = {1476-4687}
}

@article{shenMagnetoelectricTorquePolar2026,
	title = {{M}agnetoelectric {{Torque}} in {{Polar Magnetic Bilayers}}},
	author = {Shen, Zhong and Chen, Jun and Yao, Xiaoyan and Dong, Shuai},
	year = 2026,
	month = jan,
	journal = {Phys. Rev. Lett.},
	volume = {136},
	number = {1},
	pages = {016702},
	publisher = {American Physical Society},
	urldate = {2026-09-24}
}

@article{xieManipulationTopologyElectric2025,
	title = {{M}anipulation of {{Topology}} by {{Electric Field}} in {{Breathing Kagome Lattice}}},
	author = {Xie, Yu and Ji, Ke and He, Jun and Shen, Xiaofan and Wang, Dinghui and Zhang, Junting},
	year = 2025,
	month = jul,
	journal = {Phys. Rev. Lett.},
	volume = {135},
	number = {5},
	pages = {056701},
	issn = {0031-9007, 1079-7114},
	urldate = {2026-09-24},
	langid = {english}
}
\end{document}